%% file: Tortora.tex
\documentclass[natbib,graybox]{svmult}

\usepackage{mathptmx}       % selects Times Roman as basic font
\usepackage{helvet}         % selects Helvetica as sans-serif font
\usepackage{courier}        % selects Courier as typewriter font
\usepackage{type1cm}        % activate if the above 3 fonts are
\usepackage{makeidx}         % allows index generation
\usepackage{graphicx}        % standard LaTeX graphics tool
\usepackage{multicol}        % used for the two-column index
\usepackage[bottom]{footmisc}% places footnotes at page bottom

\usepackage{aas_macros}

\usepackage{floatrow}

\makeindex             % used for the subject index
\def\mst{\mbox{$M_{\star}$}}

\def\Re{\mbox{$R_{\rm e}$}}
\def\Msun{\mbox{$M_\odot$}}

\def\atlas3d{ATLAS$^{\rm 3D}$}
\def\lsim{\mathrel{\rlap{\lower3.5pt\hbox{\hskip0.5pt$\sim$}}
    \raise0.5pt\hbox{$<$}}}                % less than or approx. symbol
\def\gsim{~\rlap{$>$}{\lower 1.0ex\hbox{$\sim$}}}

\def\SN{\mbox{$S/N$}}
\def\zspec{\mbox{$z_{\rm spec}$}}
\def\zphot{\mbox{$z_{\rm phot}$}}
\def\Mauto{\mbox{{\tt MAG\_AUTO}}}
\def\Mautor{\mbox{{\tt MAG\_AUTO\_r}}}
\def\MErrautor{\mbox{{\tt MAGERR\_AUTO\_r}}}

\begin{document}

\title*{Galaxy evolution within the Kilo-Degree Survey}

\author{C.~Tortora, N.~R.~Napolitano, F.~La~Barbera, N.~Roy, M.~Radovich, F.~Getman, M.~Brescia, S.~Cavuoti, M.~Capaccioli, G.~Longo and the KiDS collaboration}
\institute{C.~Tortora \email{ctortora@na.astro.it},
N.R.~Napolitano, F.~La~Barbera, F.~Getman, M.~Brescia, S.~Cavuoti
\at INAF -- Osservatorio Astronomico di Capodimonte, Salita
Moiariello, 16, 80131 - Napoli, Italy, \and M.~Radovich \at NAF -
Osservatorio Astronomico di Padova, vicolo dell'Osservatorio 5,
I-35122 Padova, Italy \and N.~Roy, M.~Capaccioli,  G.~Longo \at
Department of Physics, University of Napoli "Federico II", via
Cinthia 9, 80126 Napoli, Italy \and KiDS collaboration:
http://kids.strw.leidenuniv.nl/team.php}

\titlerunning{Galaxy evolution within KiDS}
\authorrunning{Tortora C. et al.}

\maketitle

\abstract{The ESO Public Kilo-Degree Survey (KiDS) is an optical
wide-field imaging survey carried out with the VLT Survey
Telescope and the OmegaCAM camera. KiDS will scan 1500 square
degrees in four optical filters (u, g, r, i). Designed to be a
weak lensing survey, it is ideal for galaxy evolution studies,
thanks to the high spatial resolution of VST, the good seeing and
the photometric depth. The surface photometry have provided with
structural parameters (e.g. size and S\'ersic index), aperture and
total magnitudes have been used to derive photometric redshifts
from Machine learning methods and stellar masses/luminositites
from stellar population synthesis. Our project aimed at
investigating the evolution of the colour and structural
properties of galaxies with mass and environment up to redshift $z
\sim 0.5$ and more, to put constraints on galaxy evolution
processes, as galaxy mergers.}

\vspace{1cm}

We start from the KiDS-Data Release 2 (DR2) multi-band source
catalogs, based on source detection in the r-band images.
Star/galaxy separation is based on the {\tt CLASS\_STAR} (star
classification) and \SN\ (signal-to- noise ratio) parameters
provided by S-Extractor \cite{Bertin_Arnouts96_SEx,
LaBarbera_08_2DPHOT}. S-Extractor will provide with aperture
photometry and Kron-like magnitude \Mauto. From the original
catalog of $\sim 22$ millions of sources, the star/galaxy
separation leaves with a sample of $\sim 7$ millions of galaxies.
To perform galaxy evolution studies and determine reliable
structural parameters, the highest-quality sources have to be
taken \cite{LaBarbera_08_2DPHOT, SPIDER-I}. Thus, we have finally
selected those systems with the highest r-band signal-to-noise
ratio (\SN), $\SN \equiv$ 1/\MErrautor $> 50$, where \MErrautor\
is the uncertainty of \Mauto . This sample will consist of $\sim
380000$ galaxies. In addition to aperture photometry and Kron-like
magnitudes, the relevant data for each galaxy are listed in the
following.

\begin{figure}[h]
\floatbox[{\capbeside\thisfloatsetup{capbesideposition={left,top},capbesidewidth=4cm}}]{figure}[\FBwidth]
{\caption{2D fit output for an example galaxy from 2DPHOT. The top
panels show the galaxy image (left) and the residual after the fit
(right), while the six bottom panels provide residuals after
subtraction, plotted as a function of the distance to the galaxy
center, with each panel corresponding to a different bin of the
polar angle. Residuals are normalized with the noise expected from
the model in each pixel.}\label{fig:1}}
{\includegraphics[width=4.5cm]{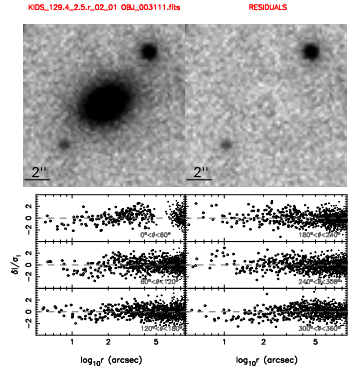}}
\end{figure}

{\it Structural parameters.} Surface photometry is performed using
the 2DPHOT environment. For each galaxy, a local PSF model is
constructed by fitting the four closest stars to that galaxy.
Galaxy images were fitted with PSF-convolved S\'ersic models
having elliptical isophotes plus a local background value
\cite{LaBarbera_08_2DPHOT}. The fit provides the following
parameters: surface brightness $\mu_{\rm e}$, effective radius,
\Re, S\'ersic index, $n$, total magnitude, $m_{S}$, axis ratio,
$q$, etc. Fig. \ref{fig:1} shows the 2DPHOT output for an example
galaxy. For further details, see the contribution from Roy et al.

{\it Photometric redshifts.} Photo-z's are derived from KiDS ugri
photometry using a supervised machine learning model, the Multi
Layer Perceptron with Quasi Newton Algorithm (MLPQNA,
\cite{Brescia+14}) within the DAMEWARE (DAta Mining and
Exploration Web Application REsource, \cite{Brescia+14_DAMEWARE}).
Supervised methods use a knowledge base (in our case with
spectroscopic redshifts) of objects for which the output (in our
case the redshift) is known a-priori to learn the mapping function
that transforms the input data (our optical magnitudes) into the
desired output (i.e. the photometric redshift). For this reason,
our sample is cross-matched with spectroscopic samples from SDSS
\cite{Ahn+14_SDSS_DR10} and GAMA \cite{Driver+11_GAMA} surveys,
which provide a redshift coverage up to $z \sim 0.8$. The
knowledge base consists of $\sim 60000$ objects. 60\% of these
objects are used as train sample, and the remaining ones for the
blind test set. The redshifts in the blind test sample resemble
the spectroscopic redshifts quite well, with a scatter in the
quantity $\vert \zspec - \zphot \vert/(1+\zspec)$ of $\sim 0.03$.
This approach reaches high accuracies with optical band only, and
is far better than traditional spectral energy distribution
(SED)-fitting methods. After these experiments the final catalogue
of redshifts for our high-\SN\ sample is produced. See Cavuoti et
al. (2015, in prep.) for  details.

{\it Rest-frame luminosities and stellar masses.} Luminosities and
stellar masses are made using the software {\tt Le Phare}
\cite{Ilbert+06}, which performs a simple $\chi^{2}$ fitting
method between the stellar population synthesis (SPS) theoretical
models and data. Models from Bruzual \& Charlot \cite{BC03} and
with a Chabrier IMF \cite{Chabrier01} are used. We adopt the
observed ugri magnitudes (and related $1\, \sigma$ uncertainties)
within a $3''$ aperture, which are corrected for Galactic
extinction. Total S\'ersic magnitudes, $m_{S}$, are used to
correct the outcomes of {\tt Le Phare} for missing flux.

\begin{figure}[h!]
%\sidecaption
\centering
\includegraphics[scale=.4]{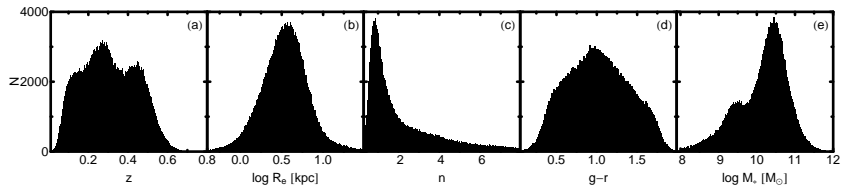}
\caption{Distribution of some galaxy parameters: photo-z, z (panel
a), r-band effective radius $\log \Re / \rm kpc$ (panel b), r-band
S\'ersic index, n (panel c), observer-frame g-r colour within an
aperture of $3''$ of radius (panel d) and Chabrier IMF-based
stellar mass $\log \mst/\Msun$ (panel e).} \label{fig:2}
\end{figure}

This high-\SN\ sample is $90\%$ complete down to a magnitude of
$\Mautor$ $\sim 21$, which means that the sample is complete at
masses $\gsim 5 \times 10^{10}$ up to redshift $z = 0.5$ (see
Napolitano et al. 2015, in prep., for further details). The
distribution of galaxy parameters for galaxies brighter than the
completeness limit are shown in Fig. \ref{fig:2}. The average
redshift of the sample is $z=0.3$, and the distribution reach the
maximum redshifts of $z=0.7$ (panel a). The distribution of
effective radii is centered on $\sim 3.5 \, \rm kpc$ average value
(panel b), while the average S\'ersic index is $\sim 1.5$, but the
distribution present a peak at $n \sim 0.8$, with a long tail to
higher values ($> 10$, panel c). Thus, our datasample is dominated
by late-type systems, which are characterized by low S\'ersic
indices \cite{Kauffmann+03, Tortora+10CG}. In the panel (d) we
show the distribution of the observer-frame g-r colour, which have
a median value of $g-r =1$ and a symmetric distribution. These
colours, when converted to rest-frame values, confirm the
considerations made looking at the distribution of S\'ersic index,
since the median rest-frame colour is $\sim 0.5$, which is typical
of a late-type blue galaxy. Finally, in panel (e) we plot the
stellar mass distribution from SPS fitting, which presents a
double-peaked distribution, with maxima at $\sim 2.5 \times
10^{9}$ and $\sim 2.5 \times 10^{10} \, \rm \Msun$, the median
mass of the sample is $1.9 \times 10^{10}\, \rm \Msun$.

{\bf Aims of the project.} The present galaxy evolution project
will aim to achieve different objectives: the analysis of the
colour-magnitude evolution, colour gradients, as the size and mass
accretion in late- and early-type systems. By the end of the
survey, we plan to collect about $3.5$ millions of galaxies with
high-quality photometry, i.e. {\it the largest sample of galaxies
with measured structural parameters in the u, g, r and i bands, up
to redshift $z=0.7$}. KiDS will provide the bridge to connect the
local environment to the high-redshift Universe, by means of this
strong characterization of the internal structure of the galaxies.
The study of the mass, colour and structural evolution of
galaxies, connected with the outcomes of theoretical models and
cosmological simulations provide a variegate set of information
about the main physical processes which shape the galaxy
evolution. In particular, taking benefit from the good pixel scale
and seeing, together with depth of VST telescope, we have scanned
the first 156 sq. deg. of KiDS searching for massive ($\mst > 8
\times 10^{10}\, \rm \Msun$) and compact ($\Re < 1.5 \, \rm kpc$)
galaxies, relic of massive galaxies at redshift $z=2$ (see Tortora
et al. 2015, in prep.). In some theoretical models the fraction of
such massive and small objects, that could survive without having
any significant transformation since $z \sim 2$, could reach a
fraction about $1-10\%$ \cite{Quilis_Trujillo13}. Previous
observational works have not detected such old, compact and
massive galaxies in the local universe
\cite{Trujillo+09_superdense}. In contrast, such peculiar objects
are found at larger redshift \cite{Damjanov+14_compacts}. A
front-edge survey like KiDS will allow us to perform the census of
these compact objects and trace their abundance and evolution in
the last billions of years, contrasting the results with
expectations from cosmological simulations. Finally, for a sample
of KiDS galaxies with SDSS spectroscopy, we will perform the Jeans
dynamical analysis of the measured velocity dispersions, to study
dark matter fraction and Initial mass function in terms of
redshift, mass, mass density and environment
\cite{TRN13_SPIDER_IMF, Tortora+14_DMslope, Tortora+14_DMevol}.
Testing the dependence of the Initial mass function by the mass
density, using dynamics and spectral indices, can help to
understand the results in Smith et al. \cite{Smith+15}, which find
shallow IMF slopes in some high-velocity dispersion galaxies, in
contrast with the current literature \cite{TRN13_SPIDER_IMF,
Spiniello+15}.

\input{referenc_talk}

\end{document}

%% file: referenc_talk.tex
%%%%%%%%%%%%%%%%%%%%%%%% referenc.tex %%%%%%%%%%%%%%%%%%%%%%%%%%%%%%
% sample references
% %
% Use this file as a template for your own input.
%
%%%%%%%%%%%%%%%%%%%%%%%% Springer-Verlag %%%%%%%%%%%%%%%%%%%%%%%%%%
%
% BibTeX users please use
% \bibliographystyle{}
% \bibliography{}
%

%mn2e
%
%science-journal

%abbrv

\bibliographystyle{plain}   % (uses fill "plain.bst")
%\bibliography{C:/Users/crescenzo/Documents/latex/Bibtex/myrefs,C:/Users/crescenzo/Documents/latex/Bibtex/myrefs_KiDS_add}